\documentclass{arxiv_preprint}

\usepackage[utf8]{inputenc}
\usepackage{amsfonts}
\usepackage{optidef}
\usepackage[ruled,vlined]{algorithm2e}
\SetAlgoCaptionSeparator{}
\journal{osajournal}
\begin{document}

\title{Constraining continuous topology optimizations to discrete solutions for photonic applications}
\author{Conner Ballew, Gregory Roberts, Tianzhe Zheng, Andrei Faraon$^{*}$}

\address{Kavli Nanoscience Institute and Thomas J. Watson Sr. Laboratory of Applied Physics, California Institute of Technology, Pasadena, California 91125, USA}

\email{\authormark{*}faraon@caltech.edu}

\begin{abstract}
Photonic topology optimization is a technique used to find the electric permittivity distribution of a device that optimizes an electromagnetic figure-of-merit. Two common techniques are used: continuous density-based optimizations that optimize a grey-scale permittivity defined over a grid, and discrete level-set optimizations that optimize the shape of the material boundary of a device. More recently, continuous optimizations have been used to find an initial seed for a concluding level-set optimization since level-set techniques tend to benefit from a well-performing initial structure. However, continuous optimizations are not guaranteed to yield sufficient initial seeds for subsequent level-set optimizations, particularly for high-contrast structures, since they are not guaranteed to converge to solutions that resemble only two discrete materials. In this work, we present a method for constraining a continuous optimization such that it converges to a discrete solution. This is done by inserting a constrained sub-optimization at each iteration of an overall gradient-based optimization. This technique can be used purely on its own to optimize a device, or it can be used to provide a nearly discrete starting point for a level-set optimization.
\end{abstract}

\section{Introduction}

Photonic inverse design is the process of choosing a figure-of-merit (FoM) and finding an optimal photonic device that maximizes or minimizes the FoM. A subset of photonic inverse design is topology optimization, which entails altering the shape of the photonic structure (i.e. its topology) to achieve the desired performance \cite{Jensen2011}. In this case the device is characterized by its electric permittivity $\epsilon(x,y,z)$ in a 3D design region.

A useful technique for inverse-design is \textit{gradient descent}. This is the process of finding a local minimum of a function $f(x)$ by following the gradient $\nabla f$. The functions being minimized are generally non-linear, thus a computed gradient is only valid in a locally linear region of $f$. Gradient descent is therefore an iterative optimization, comprised of computing $\nabla f$, stepping $f$ in the direction of $-\alpha\nabla f$ where $\alpha$ is a sufficiently small step size, and concluding the optimization when $\nabla f$ is sufficiently close to 0. A major enabling technique for gradient-based optimization is the \textit{adjoint method} for efficiently computing a gradient. The technique has been applied to many fields (see \cite{Givoli2021} for a recent overview of the topic), including photonic design \cite{Lalau-Keraly2013}.

The gradient-descent procedure becomes increasingly complex as the dimensions of the parameter space increase and as constraints are imposed. For optimizing the topology of photonic devices, the parameter space includes the permittivity of every voxel in the design region, which can be arbitrarily large. For many photonic devices, the desired solution is constrained due to the limitations of available fabrication techniques that make only certain solutions physically realizable. Some common requirements of current fabrication techniques include:

\begin{itemize}
    \item \textit{Minimum feature size}: the size of the smallest uniform material piece. This is typically dictated by lithography, material etching, and material deposition capabilities. 
    
    \item \textit{Connectivity}: one or more materials are topologically connected.
    
    \item \textit{Binary}: the device is made of only two materials.
\end{itemize}

There are many topology optimization algorithms and many reviews summarizing them \cite{Sigmund2013}. They can be characterized as density-based optimization or shape optimization: density-based optimizations feature a density variable $\rho$ that is evaluated across a grid of points, thus describing the device in an element-wise way. Density-based optimization can be further classified as discrete, where $\rho=0$ or $\rho=1$, or continuous, where $0\leq\rho\leq1$. Shape-based optimizations instead describe the boundary of the device $\Omega$. One common technique for shape-based optimizations involves describing the boundary $\Omega$ as the level-set contour of a higher-dimensional function, called the level-set method. In the context of photonic inverse design, two methods have dominated the literature: continuous density-based optimizations, and shape optimizations using the level-set method.

Of the fabrication constraints listed above, the \textit{binary} constraint is a substantial difference between the continuous density-based and the discrete level-set approaches to photonic optimization, with the former inherently describing a non-binary device and the latter inherently describing a binarized device. While the level-set method has the advantage of describing binarized devices, a known limitation is its ability to easily nucleate holes, bridge gaps, and generally converge to shapes that are substantially different from the starting topology \cite{Sigmund2013,VanDijk2013}. A technique that is now employed in photonic optimization is to begin with a density-based optimization, then switch to a level-set method to conclude the optimization \cite{Piggott2015}. Given the limitations of the level-set method, it is important that the density-based optimization converge to a sufficiently good (high-performing and nearly binary) starting point for the level-set method. This represents a challenge for devices that feature materials with large refractive index contrast, which will be shown in Section 3 of this work.

The goal of binarizing a continuous density-based optimization is not new, and is not limited to photonic inverse-design. Commonly used methods for this task include projection filters, which encourage the device to become more binary but fall short of explicitly requiring the device become binary \cite{Sigmund1998}. More advanced methods employing gradually strengthened sigmoidal or Heaviside filters have been used to aid convergence to a binary solution \cite{Sell2017,Camayd-Munoz2020}. However, these do not guarantee that the device becomes binary, and the final thresholding operation (in which all voxels are rounded to 0 or 1) can thus incur a severe performance penalty that requires substantial level-set re-optimization to overcome.

We present a method here that reformulates the gradient-based continuous density optimization in a manner that strictly enforces that the design become fully binary, thus preventing the option of settling at a non-binary solution. We have found this technique to be particularly useful in optimizing high index contrast devices. The method uses the gradient information at each iteration to step the permittivity in a direction that maximizes the performance of the device while constraining the permittivity step to binarize the device by some amount that is greater than zero. We refer to this procedure as "sub-optimization", since it occurs at every iteration of the overall photonic optimization. An important quality of this sub-optimization is that much of the computational burden of finding the ideal constrained permittivity is eased by instead solving the Lagrange dual-problem of the original sub-optimization, which converts the original N-dimensional linear optimization (where $N$ is the number of voxels in the design region) into a 1-dimensional non-linear optimization which can be easily solved computationally.

The remainder of the paper is composed as follows: first, we formulate the sub-optimization and derive the solution of the Lagrange dual problem; then, we provide an example of an optimized spectral demultiplexer. The examples are intended to illustrate the practical usage of the method, compare its performance to a traditional steepest-ascent optimization, integrate the technique with a subsequent level-set optimization, and describe a technique to control the overall fraction of material in the final device. 

\section{Lagrange dual problem}

\subsection{Setup}

Consider a discretized 3D domain of interest $\mathcal{P}$. The domain contains a region of dielectric permittivity $\epsilon_i, i\in \mathcal{D}\subseteq \mathcal{P}$, where $\mathcal{D}$ is called \textit{design region} or \textit{device}. The objective function $f(\textbf{E},\textbf{H},\vec{\epsilon})$ depends on the electric and magnetic fields output from the device, and the permittivity of the design region $\vec{\epsilon}(x)$  with $ \epsilon_i \in [\epsilon_{min},\epsilon_{max}] \: \forall i \in \mathcal{D}$. The general goal of the inverse-design optimization is to find the optimal permittivity of the design region that maximizes $f(\textbf{E},\textbf{H},\vec{\epsilon})$, and here we wish to constrain the solutions to only permittivity distributions consisting of two materials. $\epsilon_{min}$ and $\epsilon_{max}$ are the lower and upper bounds of the permittivity, respectively, representing the two materials the final device is comprised of. The electric field $\mathbf{E}$ and magnetic field $\mathbf{H}$ inside the domain can be solved for using Maxwell's equation solvers such as the finite-difference-time-domain (FDTD). The adjoint method computes the gradient $\partial f/{\partial \vec{\epsilon}}$ at all points within the design region. 

The quantities involved in 3D simulations, such as the device permittivity $\vec{\epsilon}$ and FoM gradient $\partial f/{\partial \vec{\epsilon}}$, are often computed and stored as 3D matrices. Here, we find it convenient to flatten these into N-dimensional column vectors, where $N$ is the total number of voxels in the device region. For the remainder of this derivation, all multi-dimensional quantities are considered N-dimensional vectors.

We define a metric to quantify how binary the device is. In this case, we use a simple absolute value function to define the binarization of a single pixel, whose mean across all pixels quantifies the binarization of the device. The absolute value function is scaled and shifted such that it evaluates to 1 at $\epsilon=\epsilon_{min}$ and $\epsilon=\epsilon_{max}$, 0 at the permittivity midpoint, $\epsilon_{mid}$.

\begin{gather}
    B = \frac{1}{N}\sum_i\left|\frac{\epsilon_i - \epsilon_{mid}}{\epsilon_{max}-\epsilon_{mid}}\right|, i\in \mathcal{D} \label{eq:binarization_definition}\\
    \epsilon_{mid} = \frac{\epsilon_{min}+\epsilon_{max}}{2}
\end{gather}

If we define $\vec{b}=dB/d\vec{\epsilon}$, then the approximate change in binarization of the device given a permittivity step $\vec{\Delta\epsilon}$ is $\vec{b}\cdot\vec{\Delta\epsilon}$. This term can thus be constrained to predictably control the change in device binarization.

Direct gradient ascent would shift the device permittivity in the direction of $\partial f/{\partial \vec{\epsilon}}$. However, the gradient may not point in a direction that increases the binarization of the device, so we reformulate the optimization to \textit{improve the objective function as much as possible while enforcing that the binarization increases by some amount $\beta$}. This corresponds to solving the following constrained optimization problem, which we refer to as the sub-optimization problem from now on since it is solved at every iteration of the main photonic optimization.

\begin{maxi}
	{\vec{\Delta \epsilon}}{\vec{\frac{\partial f}{\partial \epsilon}} \cdot \vec{\Delta \epsilon}}
    {\label{eq:1}}{}
	\addConstraint{\vec{b} \cdot \vec{\Delta \epsilon}}{\geq \beta}
	\addConstraint{\Delta \epsilon_{i}}{\leq u_{i}} \quad \forall i
    \addConstraint{\Delta \epsilon_{i}}{\geq l_{i}} \quad \forall i
\end{maxi}

The vectors $\vec{u}$ and $\vec{l}$ control the maximum step size that the permittivity can undergo in a single iteration. They are also used to enforce the maximum and minimum permittivity constraints (i.e. $ \epsilon_i \in [\epsilon_{min},\epsilon_{max}] \: \forall i$) since they can be set to 0 for a permittivity voxel at $\epsilon_{min}$ or $\epsilon_{max}$. 

We convert (\ref{eq:1}) to a canonical form.

\begin{mini}
	{\vec{x}}{\vec{c} \cdot \vec{x}}
    {\label{eq:2}}{}
	\addConstraint{\beta - \vec{b} \cdot \vec{x}}{\leq 0}
	\addConstraint{x_{i} - u_{i}}{\leq 0} \quad \forall i
    \addConstraint{l_{i} - x_{i}}{\leq 0} \quad \forall i
\end{mini}

Since $\vec{x} \in \mathbb{R}^N$, there are $2N+1$ inequality constraints in \eqref{eq:2} in addition to the $N$ dimensions of the non-constrained optimization problem. For large $N$, this problem is challenging to solve numerically. Fortunately, this multi-dimensional linear optimization can be reduced to a single-dimensional non-linear optimization, regardless of the value of $N$. This is done by instead solving the \textit{Lagrange dual problem}, which in this case will have the same optimal point as \eqref{eq:2} since the property of \textit{strong duality} holds for linear optimizations \cite{boyd_vandenberghe_2004}.

\subsection{Solution}

To obtain the dual problem, we start with the \textit{Lagrangian}. This is defined as:

\begin{equation}
    {\label{eq:Lagrangian}{}}
    \begin{split}
        L(\vec{x},\vec{\lambda_1}, \vec{\lambda_2}, \nu) &= \vec{c} \cdot \vec{x} + \vec{\lambda_1} \cdot (\vec{x}-\vec{u}) + \vec{\lambda_2} \cdot (\vec{l}-\vec{x}) + \nu (\beta - \vec{b} \cdot \vec{x}) \\
        &= \vec{x} \cdot (\vec{c} + \vec{\lambda_1} - \vec{\lambda_2} - \nu \vec{b}) - \vec{\lambda_1} \cdot \vec{u} + \vec{\lambda_2} \cdot \vec{l} + \nu \beta
    \end{split}
\end{equation}

The \textit{dual function} is defined as the infimum of the Lagrangian over $\vec{x}$:

\begin{equation}
    {\label{eq:dual_fun}{}}
    \begin{split}
        g(\vec{\lambda_1}, \vec{\lambda_2}, \nu) = \inf_{\vec{x}} L(\vec{x}, \vec{\lambda_1}, \vec{\lambda_2}, \nu)
    \end{split}
\end{equation}

Since \eqref{eq:Lagrangian} is linear, the infimum is $-\infty$ if any slope is non-zero.

\begin{equation}
    {\label{eq:dual_fun_2}{}}
    g(\vec{\lambda_1}, \vec{\lambda_2}, \nu) = \left\{
        \begin{array}{ll}
            -\infty & \quad \vec{c} + \vec{\lambda_1} - \vec{\lambda_2} - \nu \vec{b} \neq \vec{0} \\
            -\vec{\lambda_1} \cdot \vec{u} + \vec{\lambda_2} \cdot \vec{l} + \nu \beta & \quad  \vec{c} + \vec{\lambda_1} - \vec{\lambda_2} - \nu \vec{b} = \vec{0}
        \end{array}
    \right.
\end{equation}

The optimization we wish to solve is called the \textit{Lagrange dual problem}:

\begin{maxi}
	{}{g(\vec{\lambda_1}, \vec{\lambda_2}, \nu)}
    {\label{eq:dual_problem}}{}
    \addConstraint{\vec{\lambda_1}}{\geq 0}
	\addConstraint{\vec{\lambda_2}}{\geq 0}
	\addConstraint{\nu}{\geq 0}
\end{maxi}

Incorporating \eqref{eq:dual_fun_2} into Eq. \ref{eq:dual_problem}, we get:

\begin{maxi}
	{}{-\vec{\lambda_1} \cdot \vec{u} + \vec{\lambda_2} \cdot \vec{l} + \nu \beta}
    {\label{eq:dual_problem_2}}{}
    \addConstraint{\vec{\lambda_1}}{\geq 0}
    \addConstraint{\vec{\lambda_2}}{\geq 0}
	\addConstraint{\nu}{\geq 0}
	\addConstraint{\vec{c} + \vec{\lambda_1} - \vec{\lambda_2} - \nu \vec{b}}{= 0}
\end{maxi}

The fourth constraint of (\ref{eq:dual_problem_2}) can be used to write the objective function and the second constraint in terms of only $\nu$ and $\lambda_1$.

\begin{maxi}
	{}{-\vec{\lambda_1} \cdot \vec{u} + (\vec{c} + \vec{\lambda_1} - \nu \vec{b}) \cdot \vec{l} + \nu \beta}
    {\label{eq:dual_problem_3}}{}
    \addConstraint{\vec{\lambda_1}}{\geq 0}
	\addConstraint{\vec{\lambda_1}}{\geq \nu \vec{b} - \vec{c}}
	\addConstraint{\nu}{\geq 0}
\end{maxi}

For any $\nu$, the optimal elements of $\vec{\lambda_1}$ are as small as possible since $\vec{u} \geq 0$ and $\vec{l} \leq 0$. The solution to $\vec{\lambda_1}$ is obtained by satisfying the constraints while minimizing $\vec{\lambda_1}$. The solution for the i-th element of $\vec{\lambda_1}$ is thus:

\begin{equation}
    {\label{eq:lambda_1_sol}{}}
    \lambda_{1,i} = \left\{
        \begin{array}{ll}
            0 & \quad \nu b_i - c_i \leq 0 \\
            \nu b_i - c_i & \quad  \nu b_i - c_i > 0
        \end{array}
        = (\nu b_i - c_i) H(\nu b_i - c_i)
    \right.
\end{equation}

where $H(x)$ is the Heaviside step function. $\vec{\lambda_2}$ is solved using the fourth constraint of (\ref{eq:dual_problem_2}).

\begin{equation}
    {\label{eq:lambda_2_sol}{}}
    \vec{\lambda_2} = \vec{c} + \vec{\lambda_1} - \nu \vec{b} = \vec{c} + (\nu \vec{b} - \vec{c})H(\nu \vec{b} - \vec{c})^T - \nu \vec{b}
\end{equation}

Here, $H(\vec{x})$ represents the Heaviside function applied element-wise to $\vec{x}$. Since the optimal $\vec{\lambda_1}$ and $\vec{\lambda_2}$ are dependent on $\nu$, the dual problem is a function of only $\nu \in \mathbb{R}$. The optimization can be reduced to a single dimension as follows.

\begin{maxi}
	{\nu}{\nu \vec{b} \cdot \vec{l} + (\vec{u}-\vec{l}) \cdot (\nu \vec{b} - \vec{c})H(\nu \vec{b} - \vec{c})^T - \nu \beta}
    {\label{eq:dual_problem_nu}}{}
	\addConstraint{\nu}{\geq 0}
\end{maxi}

This problem can be solved numerically using open-source functions. We denote the optimal dual variables as $(\nu^\star, \vec{\lambda_1^\star}, \vec{\lambda_2^\star)}$. The optimal solutions of the dual problem can be mapped back to original variable $x$ using the \textit{Karush–Kuhn–Tucker (KKT) conditions}, which are a set of conditions that hold under certain regularity conditions. Fortunately, linearity of the constraints is a sufficient condition for the KKT conditions to hold. We use the KKT condition of \textit{complementary slackness} to translate $(\nu^\star, \vec{\lambda_1^\star}, \vec{\lambda_2^\star})$ to the optimal $\vec{x}$ that we call $\vec{x^\star}$.

\begin{align}
    {\label{eq:complementary_slackness}{}}
    \vec{\lambda_1^\star}(\vec{x^\star} - \vec{u})^T = \vec{0} \\
    \vec{\lambda_2^\star}(\vec{l} - \vec{x^\star})^T = \vec{0}
\end{align}

The solution is

\begin{equation}
    {\label{eq:x_sol}{}}
    x_i^\star = \left\{
        \begin{array}{ll}
            u_i & \quad \lambda_{1,i} \neq 0 \\
            l_i & \quad \lambda_{1,i} = 0
        \end{array}
    \right.
\end{equation}

\section{Example}

This section provides example optimizations to illustrate the described techniques usage and elaborate on several key points. We study device performance relative to a basic implementation of regular gradient ascent in which the permittivity is stepped in the exact direction of the gradient. We study this for three different cases of material-to-void refractive index contrast: a low index contrast of 1.5:1, a medium index contrast of 2.5:1, and a high index contrast of 3.0:1. Next, we pair this optimization technique with a level-set optimization that thresholds the continous permittivity to a binary solution and re-optimizes the device with a level-set optimization to recover the lost performance from the threshold operation. Finally, we demonstrate how this technique can be used to control the overall ratio of material:void in the final device simply by shifting the binarization function in Eq. \ref{eq:binarization_definition}.

\begin{figure}[htbp]
\centering\includegraphics[width=12cm]{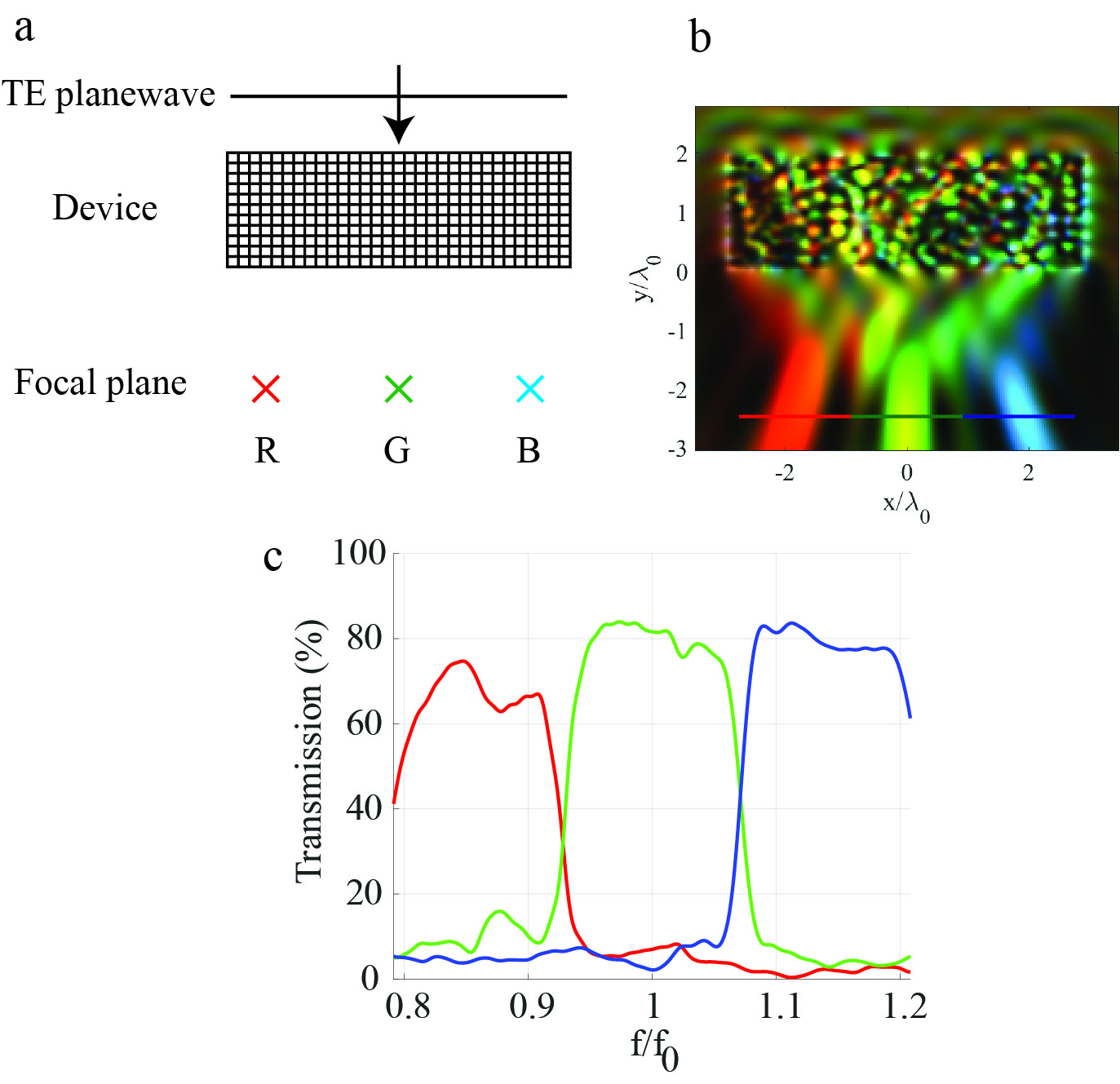}
\caption{\textbf{Example device.} (a) A schematic of the device. The device is a spectral demultiplexer that focuses three equally spaced frequency bins to three distinct points marked in the focal plane. The input is a normally incident TE-polarized planewave. (b) The intensity $|E|^2$ of each frequency overlaid. Each frequency is drawn in its equivalent hue in the visible regime. (c) The power transmission of the output fields through the apertures. Each trace is drawn in the same color as the corresponding aperture drawn in (b).}
\label{Fig1}
\end{figure}

The example optimization we use here is a spectral demultiplexer, or "color splitter". Fig. \ref{Fig1}a illustrates the purpose of the device, which focuses three equally-sized frequency bands of a normally incident TE plane-wave to three distinct points in the focal plane. The FoM function is defined as the intensity at the desired point in the focal plane, which allows us to use a simple dipole source for the adjoint source. When quantifying the performance of the device we use the power transmission through apertures centered at the dipole source. These apertures are drawn in Fig. \ref{Fig1}b as red, green, and blue horizontal lines in the focal plane. The gradient of every FoM (there is one FoM for each simulated frequency) are combined using a weighted average to obtain a single gradient vector, which is input as $\partial f/{\partial \vec{\epsilon}}$ in Eq. \ref{eq:1}. This weighting procedure is described in detail in \cite{Ballew2021}.

The device is optimized over a 42\% fractional bandwidth, which is comparable to fractional bandwidth of visible light. $\lambda_0$ and $f_0$ denote the center wavelength and frequency of the full bandwidth of the device. In Fig. \ref{Fig1}b the intensity evaluated at the various frequencies are overlaid using their equivalent hue in the visible regime. The quantitative performance of the device is shown in Fig. \ref{Fig1}c, which plots the power transmission into three monitors of size $1.56\lambda_0$, centered at the relevant focal points. Fabrication tolerance limits are not strictly enforced in this design, although the discretization of the geometry on a $\lambda_0/30$ grid and the discretization of the FDTD simulation on a $\lambda_0/10n_{max}$ grid preclude arbitrarily small features in the design. After all optimizations, we simulated the final device on a finer grid to ensure that the FDTD results had properly converged. We noticed very little change to device behavior and efficiency after halving the simulation grid size in all directions.

In the optimization problem described by Eq. \ref{eq:1}, the quantity $\beta$ describes the amount that the device must binarize after stepping the permittivity. In our implementation we set this quantity to be a factor of the max possible binarization, denoted $\beta_{max}$, during the current iteration. $\beta_{max}$ is not constant during the optimization, since points on the device eventually reach a material boundary. The maximum and minimum allowable step size for each permittivity point in the device is contained in the vectors $\vec{u}$ and $\vec{l}$, respectively, which can be used to find $\beta_{max}$. For this example, we use $\beta = 0.2\beta_{max}$. The maximum and minimum permittivity step sizes are chosen to be 0.02. The algorithm used in this example is summarized in Algorithm \ref{algorithm:1}.

\smallskip

\begin{algorithm}[H]
\label{algorithm:1}
\SetAlgoLined
    initialize device permittivity\;
    \While{$B < B_{end}$}{
        compute gradient using adjoint method, averaging multiple FoM gradients as necessary for multi-functional performance\;
        
        compute $\vec{u}$ and $\vec{l}$ based on the current permittivity and the maximum allowable step size\;
        
        compute $\beta_{max}$ based on the values of $\vec{u}$ and $\vec{l}$, and choose $\beta_0$ such that $\beta = \beta_0\beta_{max}$\;
        
        solve the optimization problem Eq. \ref{eq:dual_problem_nu} using a numerical solver\; 

        use Eq. \ref{eq:lambda_1_sol} to solve for $\lambda_1^\star$, then use Eq. \ref{eq:x_sol} to solve for $x^\star$\;
        
        update the permittivity of the device by $x^\star$\;
    }
\caption{}
\end{algorithm}

\smallskip

We compare our method with a basic implementation of gradient ascent in order to demonstrate the differences in convergence properties and final device performance. In the basic implementation of gradient ascent, we step the device permittivity in the exact direction of the electromagnetic FoM gradient (i.e. a \textit{steepest ascent} optimization). The permittivity update is computed with $\alpha (\partial f/{\partial \vec{\epsilon}})$, where $\alpha$ controls the specific step size. To ensure both stability and acceptable convergence, we dynamically alter $\alpha$ such that the maximum step in the absolute value of permittivity is 0.1. The reason for comparing these two methods is to observe their different behaviors, particularly with respect to converging to binary devices. We do not compare the speed of convergence. However, techniques for improving convergence speed (e.g. a \textit{line search} for computing an optimal step size) apply to both methods.

\begin{figure}[htbp]
\vspace*{-1.2in}
\centering\includegraphics[width=12cm]{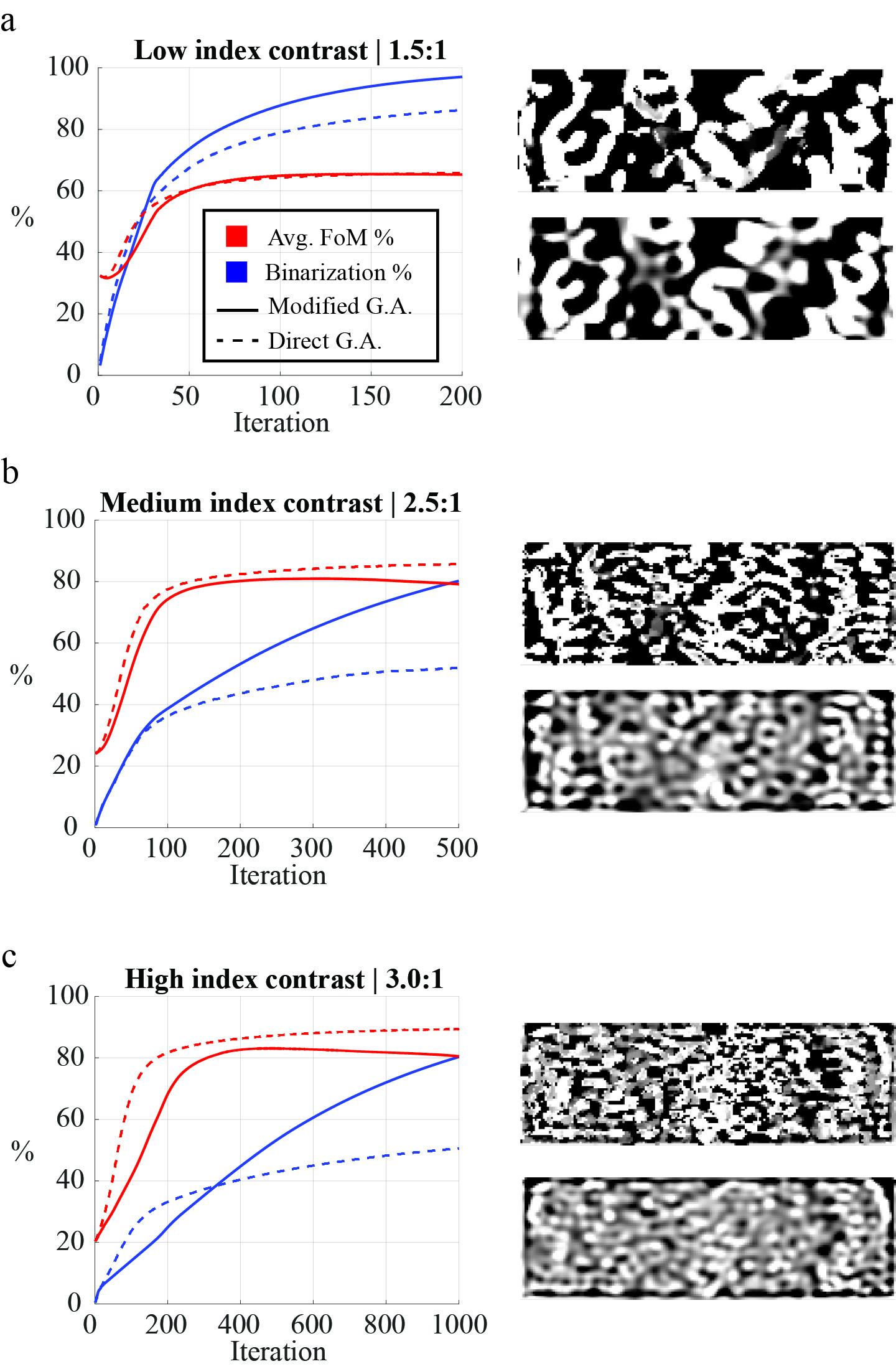}
\caption{\textbf{FoM and binarization convergence curves for different index contrasts.} The FoM (averaged across all functionalities) and binarization are drawn as red and blue traces, respectively. The modified gradient ascent (G.A.) and direct gradient ascent methods are drawn as solid and dashed lines respectively. The final index profile of each device is shown as a greyscale colormap. In each subset, the device optimized with the modified gradient ascent procedure is the top device while the device optimized with the direct gradient ascent procedure is the bottom device. The index contrasts are: (a) 1.5:1, (b) 2.5:1, (c) 3.0:1.}
\label{Fig2}
\end{figure}

The results of the traditional gradient ascent, which we call \textit{direct gradient ascent}, and the technique that this paper describes, which we call \textit{modified gradient ascent} here, are shown in Fig. \ref{Fig2} for different value of material:void refractive index contrast. The plotted average FoM is the transmission through the desired aperture averaged over all frequencies. In the low contrast case (1.5:1) in Fig. \ref{Fig2}a, the optimizations perform equally well in terms of FoM. In the medium contrast case (2.5:1) in Fig. \ref{Fig2}b, the direct gradient ascent optimization converges to a final FoM value of 84\% while the binarization stagnates at 54\%. This implies that optimization has found a local maximum of the FoM where the device is not binary. This type of solution is forbidden in the modified gradient ascent approach, which after 500 iterations is at a point where the FoM is 80\% for both the FoM and binarization (the fact that FoM and binarization percentage are nearly identical is a coincidence). In fact, the FoM reaches a maximum at iteration 300, but continues to search for a solution that is binary in order to satisfy the binarization constraint. It is necessary that the maximum FoM of a binarized device is less-than-or-equal to the maximum FoM of a greyscale device, since the former is a subset of the latter (it is possible that a binarized device exists at a better \textit{local} optimum than a seperately optimized greyscale device). The difference between the two methods becomes yet more apparent in the high index contrast case (3.0:1) in Fig. \ref{Fig2}c. In this case, the direct gradient ascent optimization converges to a 48\% binary device after 1000 iterations while the modified gradient ascent optimization reaches 80\% binarization after 1000 iterations.

\begin{figure}[htbp]
\centering\includegraphics[width=12cm]{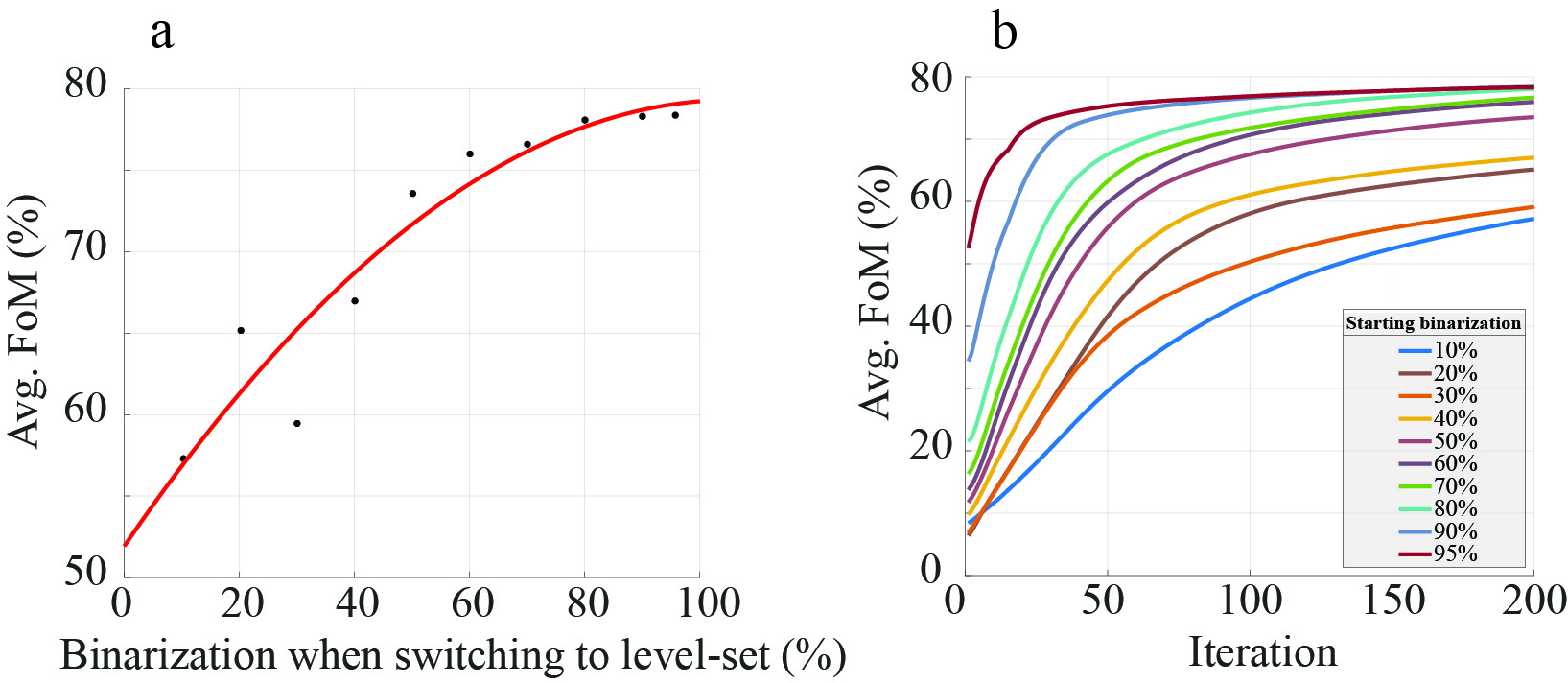}
\caption{\textbf{Combining the continuous density-based optimization with a level-set optimization.} (a) The level-set optimization uses an initial seed given by the results of the continuous density-based optimization at different starting binarization values. The average FoM is plotted as a function of this starting binarization. The red curve is a quadratic fit. (b) The convergence curves of each level-set optimization. }
\label{Fig3}
\end{figure}

While the modified gradient ascent approach described above will eventually converge to a 100\% binary device, this is not strictly required. Instead, the optimization can be concluded before the device is 100\% binary, thresholded to the nearest material boundary, and then used as an initial seed to a level-set optimization. Level-set optimizations tend to be dependent on initial seed \cite{Sigmund2013}, and we study the dependence of the converged performance of a level-set optimization with respect to the binarization percentage of an initial seed obtained with the modified gradient ascent approach. For this study, we use an implementation of level-set optimization based on signed distance functions \cite{Lebbe2019}. Fig. \ref{Fig3}a shows the converged average FoM of a 2.5:1 medium index contrast device as a function of the binarization percentage of the initial seed and Fig. \ref{Fig3}b shows the convergence curves for the individual level-set optimizations. The results suggest that the level-set optimization does indeed depend on initial seed, and that in general the more binarized starting seeds yield a better level-set optimized device. Beyond 40\% binarization, the performance monotonically increased with starting binarization, however the diminishing gains suggest that terminating the density-based optimization around 80\% will yield nearly optimal performance while minimizing the time spent in the density-based phase of the optimization.

The definition of the binarization function can be shifted with respect to permittivity in order to converge to devices with different overall material:void ratio. Eq. \ref{eq:binarization_definition} uses the mid-point of the two material permittivities as the center of the absolute value function. We can introduce a shift term, $\delta \epsilon$:

\begin{equation}
    B = \frac{1}{N}\sum_i\left|\frac{\epsilon_i + \epsilon_{mid} - \delta \epsilon}{\epsilon_{max}-\epsilon_{mid}}\right|, i\in \mathcal{D}
\end{equation}

Intuitively the binarization constraint in Eq. \ref{eq:1} can be thought of as defining whether a permittivity point should increase or decrease in order to binarize the device more, depending on its current value of permittivity. In Eq. \ref{eq:1} (with no shift about the midpoint), points that are above the midpoint are encouraged to increase further since this is the direction of binarization, and similarly points below the midpoint are encouraged to decrease further. Thus, shifting the cusp away from the midpoint affects how the permittivity is encouraged to become more binary. In general, shifting the cusp of $B$ to the left (positive $\delta \epsilon$) yields a device with more material, and vice versa. This can be a useful feature for enabling shorter maskless lithography write times or ensuring mechanical robustness without incurring a substantial decrease in device performance. 

\begin{figure}[htbp]
\centering\includegraphics[width=12cm]{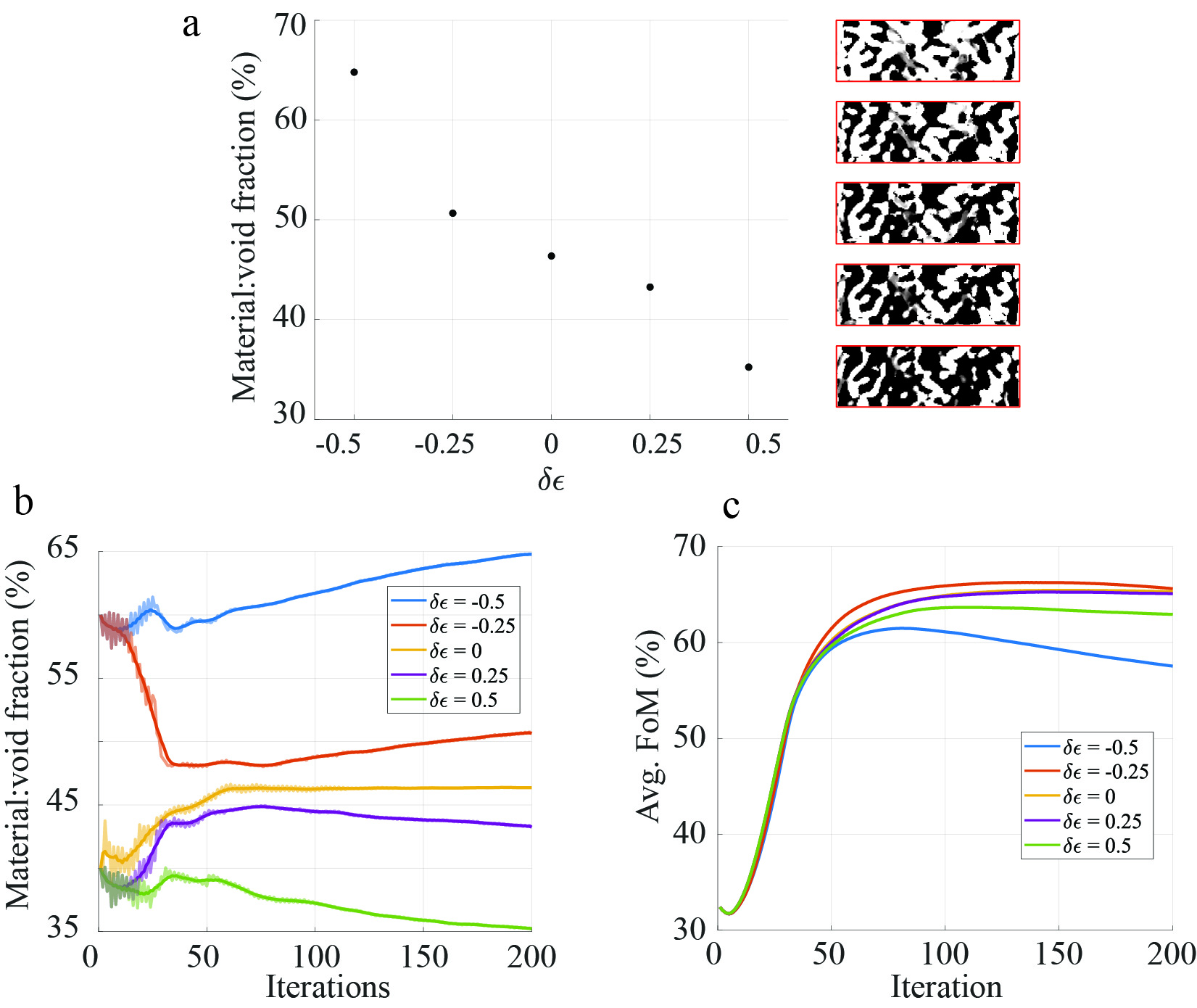}
\caption{\textbf{Controlling the overall fraction of material.} (a) The material-to-void ratio for different shifts in the definition of binarization. To the right of the plots is a picture of the resulting optimized device, where white is material and black is void. From top to bottom the devices correspond to $\delta \epsilon=-0.5$, $\delta \epsilon=-0.25$, $\delta \epsilon=0$, $\delta \epsilon=0.25$, $\delta \epsilon=0.5$. (b) The evolution of the material-to-void ratio during the optimization. Each traces shows the raw data and a 5-iteration rolling average. (c) The evolution of the FoM during the optimization.}
\label{Fig4}
\end{figure}

This technique is demonstrated in a low index contrast (1.5:1) device. Fig. \ref{Fig4}a shows the final material:void fraction as a function of $\delta \epsilon$, which is varied from -0.5 to +0.5 about the mid-point permittivity of 1.625. The resulting material distribution is also shown for each optimized device. Fig. \ref{Fig4}b shows the evolution of the material:void fraction over the course of the optimization for the different $\delta \epsilon$ values. The largest change in material:void fraction tends to occur in the early stages of optimization in this example. Fig. \ref{Fig4}c shows the evolution of the average FoM for different values of $\delta \epsilon$. The devices with the worst final performance are the ones with the most extreme $\delta \epsilon$, with the device consisting of primarily material being substantially worse than the rest. The best performing device was optimized with $\delta \epsilon = -0.25$.

\section{Conclusion}

We have presented a sub-optimization technique that is constrained to converge to fully binary solutions. The technique involves solving a sub-optimization at each iteration that maximizes the change in device performance while forcing the step in dielectric permittivity to increase the overall binarization of the device by a certain amount. This technique is computationally infeasible to solve directly with a numerical solver, but the optimization problem is reduced to a one-dimensional non-linear optimization problem by instead solving the Lagrange dual problem.

The usefulness of the technique becomes apparent when optimizing devices with a large refractive index contrast between the two material boundaries, where traditional gradient-based techniques tend to stagnate at non-binary solutions. It is shown that our level-set optimization depends on the initial seed, as has been discussed in literature over the years, and that beginning the level-set optimization with more binary seeds leads to better final device performance. Lastly, we showed an example where the material-to-void fraction of the optimized device can be controlled by simply shifting the definition of binarization. 

The specific sub-optimization presented here was intended to constrain the device only to binary solutions. However, there were no assumptions made on the binarization function $B$ other than the existence of a derivative with respect to $\epsilon$. Thus, this function can describe other properties of the device and the optimization procedure can be used to constrain other device properties in future works.

\begin{backmatter}
\bmsection{Acknowledgments}
This work was supported by the Defense Advanced Research Projects Agency EXTREME program (HR00111720035), the Jet Propulsion Laboratory, California Institute of Technology, Pasadena, CA, USA, under a contract with the National Aeronautics and Space Administration (APRA 399131.02.06.04.75 and PDRDF 107614-19AW0079), the National Institutes of Health (NIH) brain initiative program grant NIH 1R21EY029460-01, and Caltech Rothenberg Innovation Initiative.

The authors thank Oscar P. Bruno for helpful discussion during the course of this work.
\end{backmatter}

\bibliography{refs.bib}

\end{document}